\documentclass[aps,prl,reprint,twocolumn,superscriptaddress, floatfix,preprintnumbers]{revtex4-2}
\usepackage{graphicx,amsmath,amsfonts,amssymb,amsthm,xr, physics}
\usepackage{color,dsfont,upgreek}
\usepackage{mathrsfs}
\usepackage{mathtools}
\usepackage{bbold}
\usepackage{slashed}
\usepackage{float}
\usepackage[caption=false]{subfig}
\usepackage[dvipsnames]{xcolor}
\usepackage[bookmarks=true,colorlinks,linkcolor=OrangeRed,urlcolor=NavyBlue,citecolor=RoyalBlue]{hyperref}
\usepackage[capitalize]{cleveref}
\usepackage{orcidlink}

\usepackage[normalem]{ulem}

\usepackage{graphicx}
\usepackage{dcolumn}
\usepackage{bm}

\usepackage{lineno}

\begin{document}

\preprint{TUM-EFT 209/26}

\title{Schwinger Model with a Dynamical Axion}

\author{Gabriel Rouxinol}
\thanks{These authors contributed equally to this work.}
\affiliation{Department of Physics and Arnold Sommerfeld Center for Theoretical Physics (ASC), Ludwig Maximilian University of Munich, 80333 Munich, Germany}
\affiliation{Munich Center for Quantum Science and Technology (MCQST), 80799 Munich, Germany}

\author{Tom Magorsch${}^{\orcidlink{0000-0003-3890-0066}}$}
\thanks{These authors contributed equally to this work.}
\affiliation{Technical University of Munich, TUM School of Natural Sciences, Physics Department, 85748 Garching, Germany}

\author{Jesse J.~Osborne${}^{\orcidlink{0000-0003-0415-0690}}$}
\affiliation{Max Planck Institute of Quantum Optics, 85748 Garching, Germany}
\affiliation{Department of Physics and Arnold Sommerfeld Center for Theoretical Physics (ASC), Ludwig Maximilian University of Munich, 80333 Munich, Germany}
\affiliation{Munich Center for Quantum Science and Technology (MCQST), 80799 Munich, Germany}

\author{Nora Brambilla${}^{\orcidlink{0000-0003-1258-6179}}$}
\affiliation{Technical University of Munich, TUM School of Natural Sciences, Physics Department, 85748 Garching, Germany}
\affiliation{Institute for Advanced Study, Technische Universit\"at M\"unchen, 85748 Garching, Germany}
\affiliation{Munich Data Science Institute, Technische Universit\"at M\"unchen, 85748 Garching, Germany}

\author{Jad C.~Halimeh${}^{\orcidlink{0000-0002-0659-7990}}$}
\email{jad.halimeh@lmu.de}
\affiliation{Department of Physics and Arnold Sommerfeld Center for Theoretical Physics (ASC), Ludwig Maximilian University of Munich, 80333 Munich, Germany}
\affiliation{Max Planck Institute of Quantum Optics, 85748 Garching, Germany}
\affiliation{Munich Center for Quantum Science and Technology (MCQST), 80799 Munich, Germany}
\affiliation{Department of Physics, College of Science, Kyung Hee University, Seoul 02447, Republic of Korea}

\date{\today}

\begin{abstract}
One of the major open puzzles in the Standard Model of particle physics is the strong CP problem: although Quantum Chromodynamics allows a CP-violating topological $\theta$-term, experiments constrain its value to be extremely small. The Peccei–Quinn mechanism resolves this problem by promoting the $\theta$-angle to a dynamical field—introducing the axion—whose dynamics relax the effective angle $\theta_\text{eff}$ to a CP-conserving minimum. Here, we investigate the resulting axion physics in a Hamiltonian lattice gauge theory (LGT) by coupling a quantized axion field to the massive Schwinger model with a topological $\theta$-term. Using infinite matrix product state techniques, we compute the ground-state properties of the resulting theory and demonstrate that the axion dynamically relaxes $\theta_\text{eff}$ to the minimum of the vacuum energy. Consequently, the ground-state energy becomes independent of $\theta$, demonstrating the axion-mediated solution to the strong CP problem within a fully dynamical LGT. We further analyze CP restoration and extract the axion mass from the topological susceptibility and excitation spectrum. Our results provide a nonperturbative demonstration of axion dynamics in a quantum LGT amenable to investigation on modern quantum hardware.
\end{abstract}

\maketitle

\textbf{\textit{Introduction.---}}
The Standard Model of Particle Physics is the fundamental theory that describes how elementary constituents interact and build matter \cite{Weinberg1995QuantumTheoryFields, Weinberg:2004kv}. One major open puzzle in the Standard Model is the strong CP problem \cite{peskin2018introduction,schwartz2014quantum,srednicki2007quantum}. 
It arises from the fact that Quantum Chromodynamics (QCD) permits a CP-violating topological $\theta$-term, which is, however, experimentally constrained to be extremely small, requiring a fine-tuning of the parameters of the theory~\cite{kimAxionsStrong$CP$2010,Peccei:2006as,Weinberg1978NewLightBoson,Wilczek1978problem}. A solution to this problem is the Peccei--Quinn mechanism~\cite{peccei_constraints_1977,peccei$mathrmCP$ConservationPresence1977}, which postulates a scalar field with a global U$(1)_\text{PQ}$-symmetry, which is spontaneously broken, giving rise to a pseudo-Nambu--Goldstone boson: the axion. The axion couples to the topological term, effectively promoting the $\theta$-angle to a dynamical field $\theta_\text{eff}(x)=\theta+\frac{a(x)}{f_a}$, where $a(x)$ is the axion field and $f_a$ is the axion decay constant. The nonperturbative gauge dynamics generate a potential for $\theta_\text{eff}$ with a CP-conserving minimum at $\theta_\text{eff}=0\mod 2\pi$, leaving physical observables in the ground state independent of $\theta$ and thus resolving the strong CP problem~\cite{dicortonaQCDAxionPrecisely2016}.

\begin{figure}[ht]
    \centering
    \includegraphics[width=1.0\linewidth]{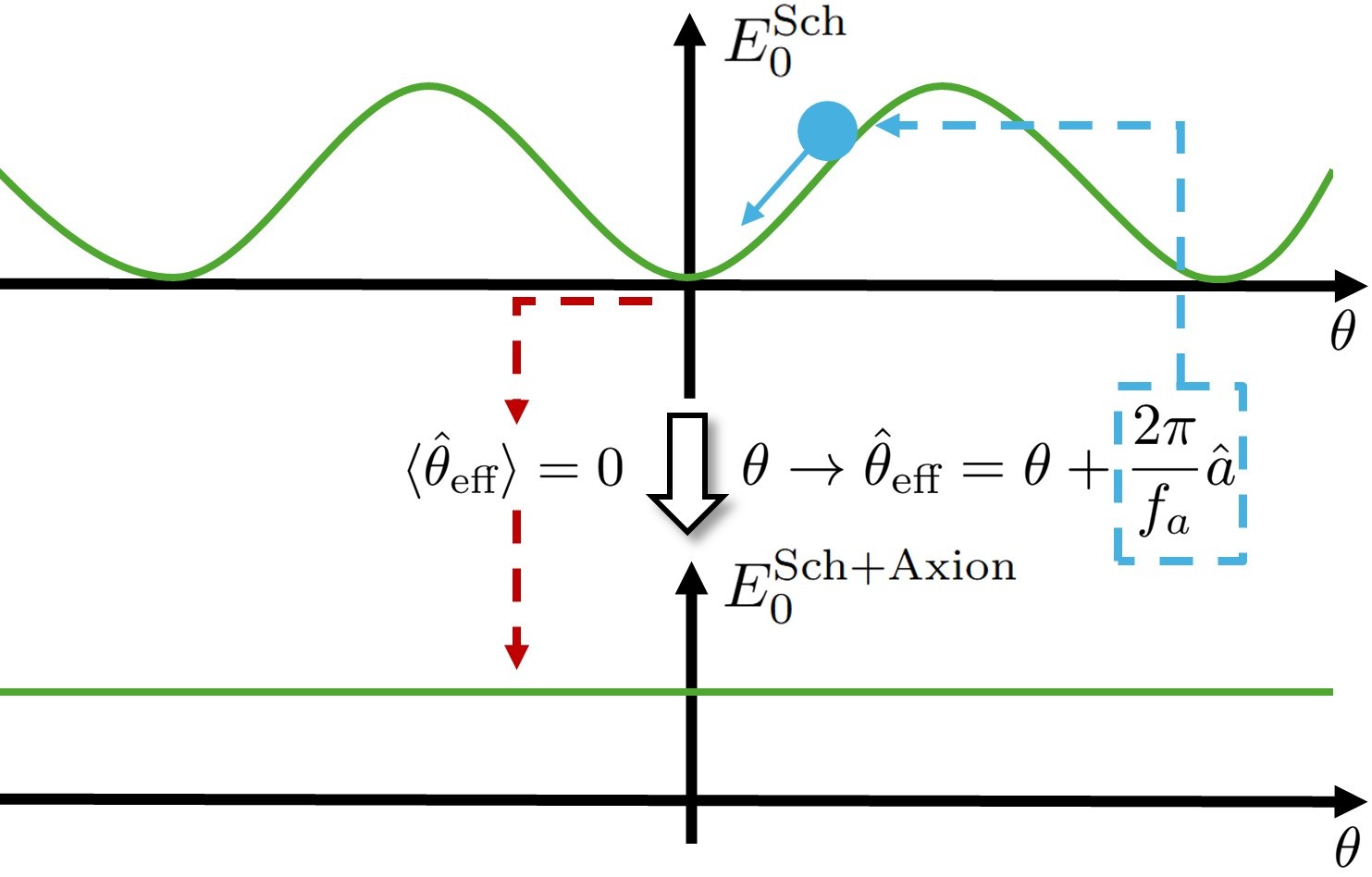}
    \caption{Schematic illustration of the effect on the ground-state energy of the Schwinger model $E_0^\text{Sch}$ induced by adding an axion field. Initially, the system is described by Hamiltonian~\eqref{eqn:ThethaSchwingerModel}, whose ground-state energy depends on $\theta$. By promoting $\theta$ to a dynamical field $\hat{\theta}_\text{eff}$, which introduces the axion field, the expected value $\langle \hat{\theta}_\text{eff} \rangle$ is driven to the minimum of the ground-state energy of the Schwinger model. This results in the ground-state energy of the Hamiltonian~\eqref{eqn:FullAxionHamiltoinan} being independent of $\theta$. }
    \label{fig:Scheme}
\end{figure}

The properties of the axion have been studied extensively, since, beyond resolving the strong CP problem, the axion is a well-motivated dark matter candidate~\cite{Preskill:1982cy,Marsh:2015xka}. However, a full characterization of the axion potential from first-principles lattice QCD calculations is challenging, as the $\theta$-term leads to a complex phase in the Euclidean path integral and hence a sign problem~\cite{Vicari:2008jw}.
A complementary nonperturbative approach is the Hamiltonian formulation~\cite{kogutHamiltonianFormulationWilsons1975}, which permits direct computations at finite $\theta$ and allows the inclusion of the axion as an explicit degree of freedom to study its interplay with the gauge fields. Furthermore, the Hamiltonian formulation of LGTs opens the door to investigations of dynamical axion physics on modern quantum hardware as part of the ongoing effort of quantum computing high-energy physics (HEP)~\cite{Byrnes2006SimulatingLatticeGauge, Dalmonte2016LatticeGaugeTheory, Zohar2015QuantumSimulationsLattice, Aidelsburger:2021mia, Zohar2021QuantumSimulationLattice, 
Barata2022MediumInducedJetBroadening,Klco2022StandardModelPhysics,Barata2023QuantumSimulationInMediumQCDJets,Barata2023RealTimeDynamicsofHyperonSpin, Bauer2023QuantumSimulationHighEnergy, Bauer2023QuantumSimulationFundamental,
DiMeglio2024QuantumComputingHighEnergy, Cheng2024EmergentGaugeTheory, Halimeh2022StabilizingGaugeTheories, Cohen2021QuantumAlgorithmsTransport,Barata2025ProbingCelestialEnergy, Lee2025QuantumComputingEnergy, Turro2024ClassicalQuantumComputing,Halimeh2023ColdatomQuantumSimulators,Bauer2025EfficientUseQuantum,Halimeh2025QuantumSimulationOutofequilibrium}, complementing a large body of quantum simulation experiments in the field~\cite{Martinez2016RealtimeDynamicsLattice, Klco2018QuantumclassicalComputationSchwinger,Gorg2019RealizationDensitydependentPeierls, Schweizer2019FloquetApproachZ2, Mil2020ScalableRealizationLocal, Yang2020ObservationGaugeInvariance, Wang2022ObservationEmergent$mathbbZ_2$, Su2023ObservationManybodyScarring, Zhou2022ThermalizationDynamicsGauge, Wang2023InterrelatedThermalizationQuantum, Zhang2025ObservationMicroscopicConfinement, Zhu2024ProbingFalseVacuum, Ciavarella2021TrailheadQuantumSimulation, Ciavarella2022PreparationSU3Lattice, Ciavarella2023QuantumSimulationLattice-1, Ciavarella2024QuantumSimulationSU3, 
Gustafson2024PrimitiveQuantumGates, Gustafson2024PrimitiveQuantumGates-1, Lamm2024BlockEncodingsDiscrete, Farrell2023PreparationsQuantumSimulations-1, Farrell2023PreparationsQuantumSimulations, 
Farrell2024ScalableCircuitsPreparing,
Farrell2024QuantumSimulationsHadron, Li2024SequencyHierarchyTruncation, Zemlevskiy2025ScalableQuantumSimulations, Lewis2019QubitModelU1, Atas2021SU2HadronsQuantum, ARahman:2022tkr, Atas2023SimulatingOnedimensionalQuantum, Mendicelli2023RealTimeEvolution, Kavaki2024SquarePlaquettesTriamond, Than2024PhaseDiagramQuantum, Angelides:2023noe, Gyawali2025ObservationDisorderfreeLocalization,  
Mildenberger2025Confinement$$mathbbZ_2$$Lattice, Schuhmacher2025ObservationHadronScattering, Davoudi2025QuantumComputationHadron, Saner2025RealTimeObservationAharonovBohm, Xiang2025RealtimeScatteringFreezeout, Wang2025ObservationInelasticMeson,li2025frameworkquantumsimulationsenergyloss,mark2025observationballisticplasmamemory,froland2025simulatingfullygaugefixedsu2,Hudomal2025ErgodicityBreakingMeetsCriticality,hayata2026onsetthermalizationqdeformedsu2,Cochran2025VisualizingDynamicsCharges, Gonzalez-Cuadra2025ObservationStringBreaking, Crippa2024AnalysisConfinementString, De2024ObservationStringbreakingDynamics, Liu2024StringBreakingMechanism, Alexandrou:2025vaj,Cobos2025RealTimeDynamics2+1D}. It also facilitates probing gauge invariance-induced nonergodic quantum many-body dynamics of relevance to HEP and condensed matter~\cite{Smith2017DisorderFreeLocalization,Brenes2018ManyBodyLocalization,Smith2017AbsenceOfErgodicity,Karpov2021DisorderFreeLocalization,Sous2021PhononInducedDisorder,Chakraborty2022DisorderFreeLocalization,Halimeh2022EnhancingDisorderFreeLocalization,Surace2020LatticeGaugeTheories,Lang2022DisorderFreeLocalization,Desaules2023WeakErgodicityBreaking,Desaules2023ProminentQuantumManyBodyScars,Aramthottil2022ScarStates,Tarabunga2023ManyBodyMagic,Desaules2024ergodicitybreaking,Desaules2024MassAssistedLocalDeconfinement,Hudomal2022DrivingQuantumManyBodyScars,Jeyaretnam2025HilbertSpaceFragmentation,Smith2025Nonstabilizerness,Falcao2025nonstabilizerness,Esposito2025magicdiscretelatticegaugetheories,Ciavarella2025GenericHilbertSpaceFragmentation,Ciavarella:2025tdl,Steinegger2025GeometricFragmentationAnomalousThermalization,Ebner2024EntanglementEntropy,Halimeh2023robustquantummany,Iadecola2020QuantumManyBodyScar,Banerjee2021QuantumScarsZeroModes,Biswas2022ScarsFromProtectedZeroModes,Daniel2023BridgingQuantumCriticality,Sau2024sublatticescarsbeyond,Osborne2024QuantumManyBodyScarring,Budde2024QuantumManyBodyScars,Calajo2025QuantumManyBodyScarringNonAbelian,Hartse2025StabilizerScars,cataldi2025disorderfreelocalizationfragmentationnonabelian}.

Motivated by this, in this work we consider the lattice Schwinger model, a U$(1)$ gauge theory in $1+1$D, coupled to a quantized axion field, as a setting to study axion--$\theta$ physics. The Schwinger model exhibits an axial anomaly, a nontrivial vacuum structure, and a CP-violating topological $\theta$-term \cite{Byrnes2002densitymatrixrenormalizationgroup,Buyens2014matrixproductstates,Shimizu2014criticalbehavior,Buyens2016confinementandstringbreaking}, thereby qualitatively capturing key features relevant for the strong CP problem.

\textbf{\textit{Model.---}}
We consider the U$(1)$ gauge theory in $1+1$D coupled to a massive Dirac fermion, known as the massive Schwinger model~\cite{schwingerGaugeInvarianceMass1962, schwinger_gauge_1962, aoki1+1dimensionalQCDFundamental1995}, with the Lagrangian density $\mathcal{L} = \overline{\psi}(i \slashed{\partial} - g\slashed{A} -m)\psi - \frac{1}{4}F_{\mu\nu}F^{\mu\nu}+\frac{g\theta}{4\pi}\varepsilon^{\mu\nu}F_{\mu\nu}$, where $\psi$ is a two-component Dirac fermion of mass $m$ and $F_{\mu\nu}=\partial_\mu A_\nu -\partial_\nu A_\mu$, with $A_\mu$ the U$(1)$ gauge field and $\mu,\nu=0,1$. The term proportional to the $\theta$-angle is a consequence of the nontrivial topological structure of the vacuum, which generates a background electric field, a feature unique to the $m\neq 0$ case, as for $m=0$ it can be rotated out, having no physical impact \cite{colemanMoreMassiveSchwinger1976, colemanChargeShieldingQuark1975}. Moreover, this model can be studied analytically using bosonization, which provides access to many parameter regimes~\cite{TongGaugeTheory2018}. In $1+1$D, the $\theta$-term is C- and P-odd but CP-even. Below, we follow the common convention in the Schwinger model and refer to the discrete transformation $E\to -E$ as CP. This transformation is not literally CP in the strict field-theoretic sense, but in the Hamiltonian formulation of the Schwinger model, it plays the same role as CP does in $3+1$D QCD. In this work, we will use the lattice formulation of the Hamiltonian density in the temporal gauge $A_0=0$ given by~\cite{hamerMassiveSchwingerModel1982, gattringerLatticeSchwingerModel1996, melnikovLatticeSchwingerModel2000}
\begin{align}
\hat{H}_\text{Sch}&(\theta)
= -\frac{\kappa}{2} \sum_{\ell=1}^{L-1}
\Big( \hat{\psi}^\dagger_\ell \hat{U}_{\ell,\ell+1} \hat{\psi}_{\ell+1}
+ \text{H.c.} \Big) \label{eqn:ThethaSchwingerModel} \\
\quad& +m\sum_{\ell=1}^{L} (-1)^\ell \hat{\psi}^\dagger_\ell \hat{\psi}_\ell
+ \frac{1}{2} \sum_{\ell=1}^{L-1}
\Big[\hat{E}_{\ell,\ell+1} + E_{\text{bg}}(\theta) \Big]^2,\nonumber
\end{align}
where matter on site $\ell$ is described by the Kogut--Susskind staggered fermion formulation~\cite{kogutHamiltonianFormulationWilsons1975, zohar_formulation_2015} with ladder operators $\hat{\psi},\hat{\psi}^\dagger$ and mass $m$. The minimal coupling $\kappa=2$ and the lattice spacing is set to unity throughout. Furthermore, we use the Wilson formulation where the gauge fields live on the link that connects the sites $\ell$ and $\ell+1$ and are represented by the infinite-dimensional operators $\hat{E}_{\ell,\ell+1}$ and $\hat{U}_{\ell,\ell+1}$. The total number of lattice sites is $L$. Furthermore, $E_{\text{bg}}(\theta) = \frac{g\theta}{2\pi}$ is the background electric field generated by the presence of the $\theta$-term. The physical states $\ket{\text{phys}}$ of the theory are the ones that respect the lattice version of Gauss' law $\hat{G}_\ell\ket{\text{phys}}=0$, with $\hat{G}_\ell = \hat{E}_{\ell, \ell +1} - \hat{E}_{\ell-1, \ell} -g \left[\hat{\psi}^\dagger_\ell\hat{\psi}_\ell + \frac{(-1)^\ell -1}{2} \right]$.

In the case of QCD, the $\theta$-term makes the vacuum energy a $2\pi$-periodic function of $\theta$, with CP-conserving minima at $\theta= 0 \mod 2\pi$. To address the strong CP problem, the Peccei-Quinn mechanism introduces a dynamical axion field $a(x)$ coupling to the topological term~\cite{peccei_constraints_1977,peccei$mathrmCP$ConservationPresence1977}, effectively promoting the $\theta$ angle to $\theta_\text{eff}(x)=\theta + \frac{2\pi}{f_a} a(x)$, where we redefined the axion decay constant $f_a$ to contain a factor $2\pi$. The axion possesses a shift symmetry $a(x)\to a(x)+c$, which is broken only by the coupling to the topological term, ensuring that its vacuum energy is determined solely by the coupling to the gauge sector. The dynamical axion resolves the strong CP problem by relaxing the effective angle towards the CP-conserving minimum $\langle \theta_\text{eff}\rangle=0 \mod 2\pi$. In our model, we can understand this phenomenon in terms of the ground-state energy of the Schwinger model $E^\text{Sch}_{0}(\theta)$, which defines an effective potential $V(\theta)$ for the axion. When $\theta$ is promoted to the field $\hat{\theta}_\text{eff}=\theta+\frac{2\pi}{f_a}\hat a$, with $\hat a$ the axion operator, the effective angle dynamically relaxes towards the minimum of the potential. This leaves the ground-state energy independent of $\theta$, as schematically shown in Fig.~\ref{fig:Scheme}. Previous works that studied the axion in the Schwinger model \cite{ho2026quantumsimulationstrongchargeparity} introduced this potential as an effective description of the theory and minimized it to obtain the desired axion properties. Our work is the first study of a dynamical axion interacting with the nonperturbative potential generated by the Schwinger model.

Promoting the $\theta$ parameter to the field $\hat{\theta}_\text{eff}$ and discretizing the axion field such that for each gauge link $\hat{E}_{\ell,\ell+1}$ there is one axion $\hat{a}_\ell$ coupled to it, we obtain $\hat{H}
= \hat{H}_\text{Sch}\big(\theta+\frac{2\pi}{f_a}\hat{a}\big) + \hat{H}_\text{kin}$, with $\hat{H}_\text{kin}=\sum_{\ell=1}^{L} \Big[\frac{\hat{p}_\ell^2}{2}+\frac{(\hat{a}_{\ell+1}-\hat{a}_{\ell})^2}{2}\Big]$, the kinetic Hamiltonian of the axion, where $\hat{p}_\ell$ is the bosonic momentum operator. To make the above model numerically tractable, one must truncate the dimension of the gauge fields. One way to achieve that is through the quantum link model (QLM) formulation \cite{chandrasekharanQuantumLinkModels1997, wieseUltracoldQuantumGases2013, yangAnalogQuantumSimulation2016,Kasper2017ImplementingQuantumElectrodynamics}, where the parallel transporter $\hat{U}_{\ell,\ell+1}$ is mapped to the spin-$s$ raising operator, $\hat{s}_{\ell,\ell+1}^+$, while the electric field becomes $g\hat{s}^z_{\ell,\ell+1}$. This transformation allows us to recover the correct commutation relations between $\hat{E}_{\ell,\ell+1}$ and $\hat{U}_{\ell,\ell+1}$ in the $s\to \infty$ limit. More details can be found in \cite{halimehTuningTopological$ensuremaththeta$Angle2022,Cheng2022tunable}. The QLM Hamiltonian becomes 
\begin{align}\nonumber
\hat{H}
=& -\frac{\kappa}{2}\sum_{\ell=1}^{L-1}
\left( \hat{\psi}^\dagger_\ell \hat{s}^+_{\ell,\ell+1} \hat{\psi}_{\ell+1}
+ \text{H.c.} \right)\\ \nonumber
&
 +g^2 \sum_{\ell=1}^{L-1}
\Bigg[ \frac{(\hat{s}^z_{\ell,\ell+1})^2}{2} + \eta \hat{s}^z_{\ell,\ell+1}+ \frac{\hat{a}_\ell}{f_a}\hat{s}^z_{\ell, \ell+1} +\frac{\eta^2}{2}  \Bigg]   \\\label{eqn:FullAxionHamiltoinan}
&  +\frac{m}{2}  \sum_{\ell=1}^{L} (-1)^\ell \hat{\psi}^\dagger_\ell \hat{\psi}_\ell + \hat{H}_a,
\end{align}
where $\hat{H}_a=\sum_{\ell=1}^{L} \Big[\frac{\hat{p}_\ell^2}{2}+\frac{(\hat{a}_{\ell+1}-\hat{a}_{\ell})^2}{2}+g^2\eta\frac{\hat{a}_\ell}{f_a}+\frac{g^2}{2}\frac{\hat{a}^2_\ell}{f_a^2}\Big]$ are the pure axionic terms of the Hamiltonian composed of a contribution from the kinetic energy of the axionic field, and one term arising from the expansion of the quadratic term containing the axion field. We define $\eta=\frac{\theta}{2\pi} -\frac{1}{4}\left[1-(-1)^{2s}\right]$, which effectively implements a transformation of $\theta \to \theta -\pi$ for half-integer spins, as in this case it is known that the QLM naturally implements a topological angle of $\pi$ \cite{suraceLatticeGaugeTheories2020, halimehAchievingQuantumField2022,Zache2022TowardContinuumLimit}. 

\begin{figure}[ht]
    \centering
    \includegraphics[width=1.0\linewidth]{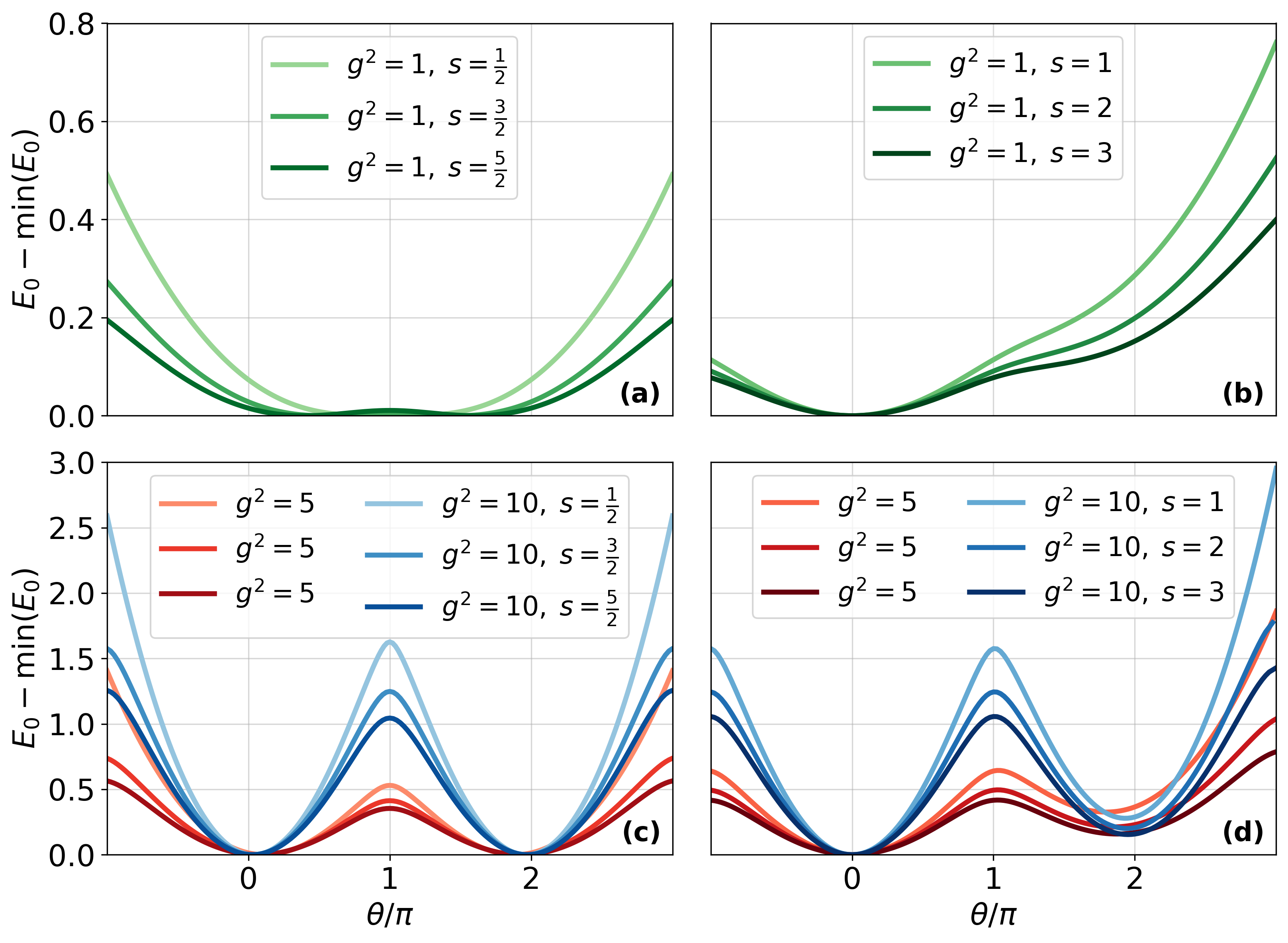}
    \caption{iDMRG calculations of the ground-state energy dependence of Hamiltonian~\eqref{eqn:FullAxionHamiltoinan} without dynamical axion fields on the topological $\theta$-angle at (a) $g^2=1$ and $s\in \{\frac{1}{2}, \frac{3}{2}, \frac{5}{2}\}$, (b) $g^2=1$ and $s\in \{1, 2, 3\}$, (c) $g^2=5, 10$ and $s\in \{\frac{1}{2}, \frac{3}{2}, \frac{5}{2}\}$, (d) $g^2=5, 10$ and $s\in \{1, 2, 3\}$, with increasing $s$ going from lighter to darker shades in the curves. For half-integer spin on plots (a) and (c), as $s$ or $g^2$ increases, the minima move closer to the expected $\theta=0,2\pi$, although for $g^2=1$ the shown truncations still lead to sizable shifts of the minima. For integer spin on plots (b) and (d), the minima are located at the expected values for $g^2=5,10$, while there is no minimum at $\theta=2\pi$ for $g^2=1$. The minima at $\theta =0, 2\pi$ do not show the desired $2\pi$ periodicity, but as $s$ increases, the shape of the axion potential moves closer to showing a period of $2\pi$.}
    \label{fig:AxionPotential}
\end{figure}

\textbf{\textit{Results.---}}
To observe the effect of a dynamical axion on the ground state of the Schwinger model, we first investigate the dependence of its ground-state energy on $\theta$ without the additional axion field. To this end, we approximate the system numerically as an infinite matrix product state (iMPS) and determine the ground state of Hamiltonian~\eqref{eqn:FullAxionHamiltoinan} using the infinite density matrix renormalization group (iDMRG) algorithm~\cite{white1992,white1993,mcculloch2008,schollwoeck2011,mptoolkit}. For the most stringent iDMRG calculations we have run, we find convergence for a bond dimension of 30 (as we are studying a system in one spatial dimension away from criticality, a small bond dimension is sufficient to represent the state accurately). All energy values presented will be the energy per unit cell consisting of a fermion site, an antifermion site, two gauge boson sites, and two axion sites. We consider the case of $\hat{\theta}_\mathrm{eff}=\theta$, dropping the axion sector, to obtain the shape of $E^\text{Sch}_0(\theta)$ in this model, which will dictate the behavior of the axion upon its introduction. We study the spin truncations $s\in \{\frac{1}{2}, 1, \frac{3}{2}, 2, \frac{5}{2}, 3\}$, with gauge coupling values of $g^2\in \{1,5,10\}$, $\kappa=2$, and $m=0.25$. The results for the ground-state energies of the Schwinger model in terms of $\theta$ are shown in Fig.~\ref{fig:AxionPotential}. We observe in Fig.~\ref{fig:AxionPotential}(a,c) that half-integer spins lead to a symmetry around $\theta=\pi$, and two global minima. For lower $g^2$ and smaller spin truncations, the minima are shifted from the expected $\theta=0$ and $\theta=2\pi$. However, they approach these values as the spin truncation increases. On the other hand, for integer spins, shown in Fig.~\ref{fig:AxionPotential}(b,d), there is a global minimum exactly at $\theta = 0$, while there is no periodicity. However, for higher $g$ there is a second, local, minimum, which moves closer to $\theta=2\pi$ and to the value of the global minimum as the gauge-field truncation increases. This behavior is a consequence of the symmetries of the Hamiltonian~\eqref{eqn:FullAxionHamiltoinan}, as we observe that in the absence of axions, the transformation $\eta \to -\eta$ and $\hat{s}^z_{\ell, \ell+1} \to -\hat{s}^z_{\ell, \ell+1}$ leaves it invariant. In the half-integer formulation, this is equivalent to transformations of $\theta \to \pi -\theta$, hence the symmetry around $\pi$, while for integer spins it stems from $\theta \to -\theta$, resulting in an even potential. It is expected that, as the truncation is increased and we approach the $s\to \infty$ limit, the well-located minima of integer spins and the periodicity of half-integer spins will together result in $2\pi$ periodicity and minima at $0\mod 2\pi$, as is the case in QCD. A more complete plot using higher electric-field truncations is shown in the Supplemental Material~\cite{SM}, further substantiating the above claim and showing that, even though we are far from the continuum limit, the model already captures the general qualitative features of the continuum theory. For lower values of $g^2$, the results are more sensitive to truncation errors, as the only term in the Hamiltonian that depends on $\theta$ is proportional to $g^2$. Consequently, for smaller $g^2$, $\hat{s}^z_{\ell, \ell +1}$ would need to take larger values to reproduce a similar mechanism, which is prohibited by the truncation, as it limits the maximum expectation value the operator can attain.

\begin{figure}[t!]
    \centering
    \includegraphics[width=1.0\linewidth]{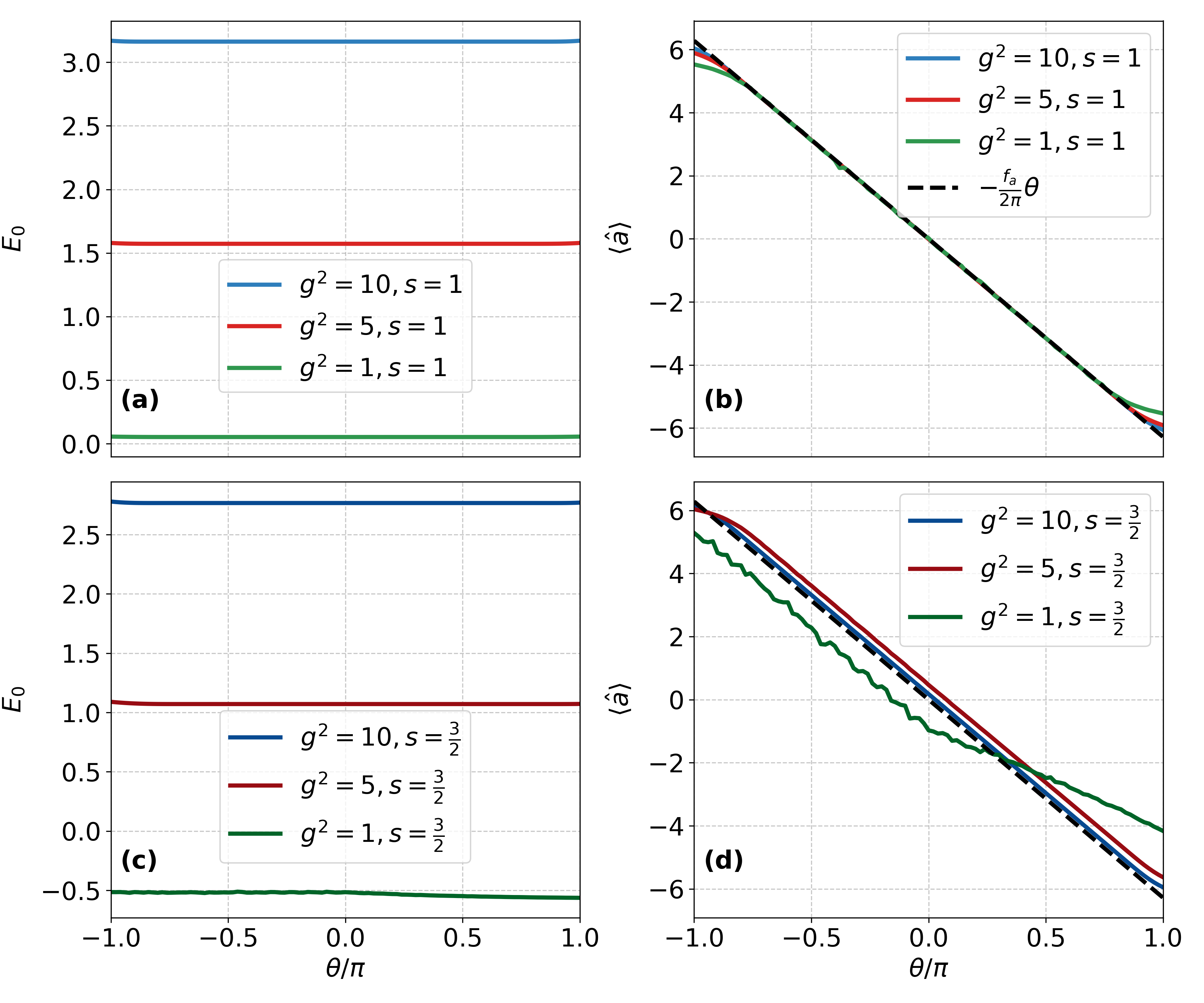}
    \caption{iDMRG calculations of the (a) dependence of the ground-state energy of Hamiltonian~\eqref{eqn:FullAxionHamiltoinan} on the topological $\theta$-angle and (b) the expectation value of the axion field $\hat{a}$ as a function of $\theta$, for the $s=1$ QLM. An equivalent study is done for the $s=\frac{3}{2}$ QLM in plots (c) and (d). In (a) and (c), the energy shows no dependence on the $\theta$ angle, as the axion drives the effective angle to the minimum of the axion potential $\theta_\text{min}$. The only exception is in (c) for $g^2=1$, where a slight deviation from a flat line is observed, due to the flatness of the axion potential in this case. On (b) and (d), the overlap between $\langle \hat{a} \rangle$ and $-\frac{f_a\theta}{2\pi}$ confirms that $\langle\hat{\theta}_\text{eff}\rangle=0$ for $s=1$. On the other hand, for $s=\frac{3}{2}$ we have $\langle\hat{\theta}_\text{eff}\rangle=\theta_\text{min}$, a constant value for $g^2=5,10$, which grows closer to zero as we increase $g^2$. Similarly, because of the axion potential's flatness, the behavior is much more irregular for $g^2=1$.}
    
    \label{fig:KineticAxionSpin3_2}
\end{figure}

Including the dynamical axion field and restricting ourselves to $\theta \in ]-\pi, \pi[$, we expect the effective angle $\hat\theta_\text{eff}$ to align with the expectation value that minimizes the ground-state energy of the Schwinger model.
We therefore consider the Hamiltonian~\eqref{eqn:FullAxionHamiltoinan} including the axion field for the same values of $g^2$ and fix $s=1, \frac{3}{2}$, with $f_a=4\pi$ and a local truncation of the axion field to $32$ levels per site. Using the same techniques and parameters as before, we obtain the ground-state energy of the full Hamiltonian with the axion field for different values of $\theta$. For $s=1$, the results for both the ground-state energy and the expectation of the axion field $\langle\hat a\rangle$ are shown in Fig.~\ref{fig:KineticAxionSpin3_2}(a,b), respectively. Here, the inclusion of the axion drives the energy to the minimum value, which, as shown previously, is obtained when $\langle \hat{\theta}_\text{eff}\rangle = \theta_\text{min}=0$. This minimization implies not only that $\langle\hat{a}_\ell\rangle=-\frac{f_a\theta}{2\pi}$, but also that the ground-state energy remains constant for all values of $\theta$. Consequently, in Fig.~\ref{fig:KineticAxionSpin3_2}(a), we find the ground-state energy to be completely flat for all values of $\theta$ and for the three chosen values of $g^2$, mirroring $3+1$D QCD, where the axion removes the $\theta$-dependence of the vacuum energy. As stated above, this effect is due to the cancellation between the axion and the $\theta$-term as shown in Fig.~\ref{fig:KineticAxionSpin3_2}(b). Effectively, here we show that for the low truncation of $s=1$, the introduction of the axion leads to $\langle\hat{\theta}_\text{eff}\rangle=0$ in the ground state, demonstrating the axion-mediated solution to the strong CP problem in our model.

We perform the same study for a truncation of $s=\frac{3}{2}$ in Fig.~\ref{fig:KineticAxionSpin3_2}(c,d). As the minima of the potential are no longer situated at zero, but at values close to it for $g^2=5,10$ and further shifted from it for $g^2=1$ (see Fig.~\ref{fig:AxionPotential}(a,c)), we expect $\langle\hat\theta_\text{eff}\rangle$ to take small but nonzero values and thus the cancellation to not work exactly. In Fig.~\ref{fig:KineticAxionSpin3_2}(c), we show the ground-state energy and observe that it is independent of the topological angle for $g^2=5,10$, while for $g^2=1$, a small dependence can be seen for positive $\theta$. In Fig.~\ref{fig:KineticAxionSpin3_2}(d), we show the expectation value of the axion field. We observe that due to the shift of the minimum from $\langle\hat\theta_\text{eff}\rangle=0$, the axion field does not exactly cancel with the $\theta$-term. This is especially noticeable for $g^2=1$, where the flatness of the potential near the minimum introduces numerical errors. However, this is expected as a consequence of the truncation of the electric field, while increasing the spin truncation will move the minimum to the correct $\theta_\text{min}=0$.

\begin{figure}[t!]
    \centering
    \includegraphics[width=1.0\linewidth]{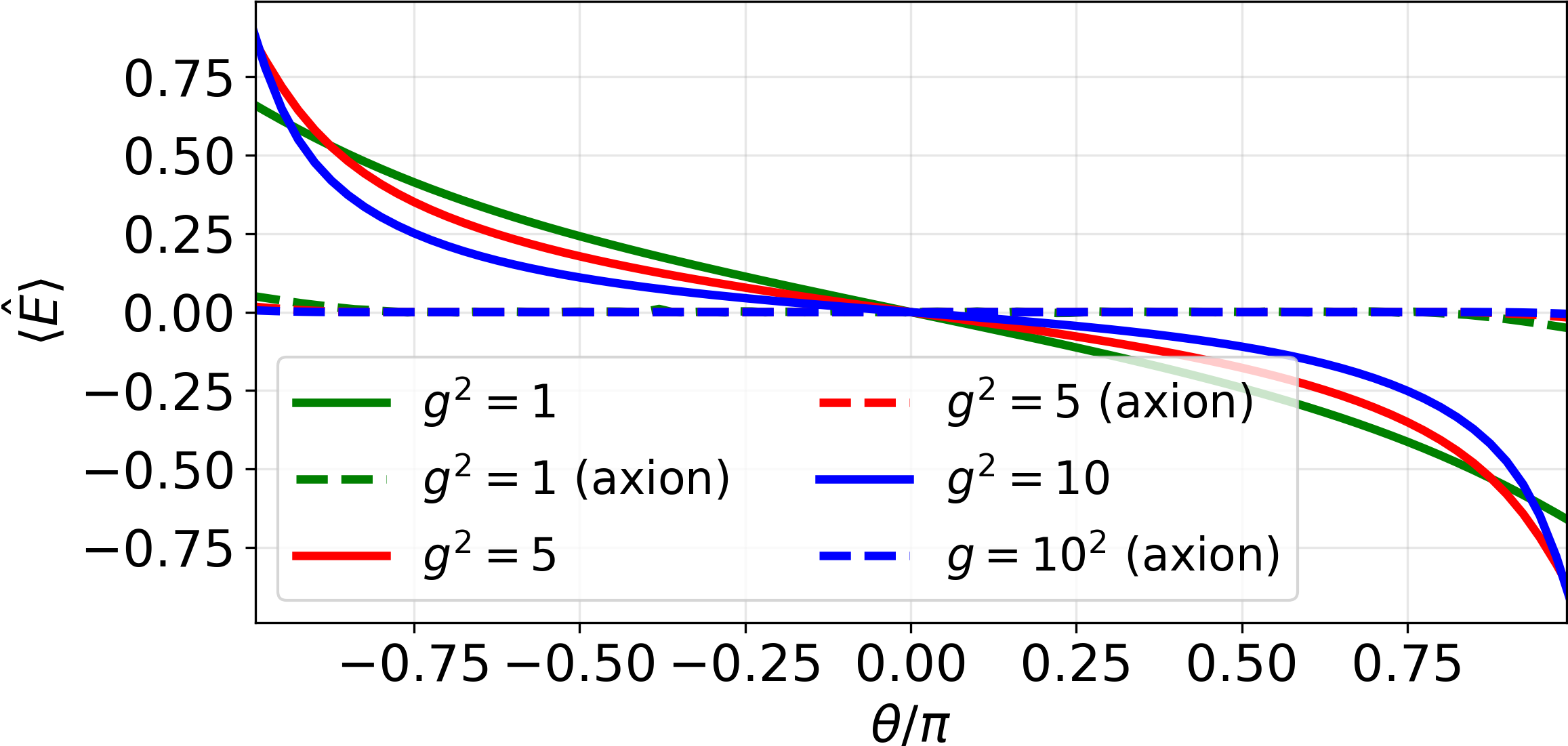}
    \caption{Expectation value of the electric field $\langle \hat{E}\rangle$ as a function of $\theta$ for the massive Schwinger model with axions (dashed lines) and without axions (solid lines) for $g^2\in\{1,5,10\}$ and a QLM truncation of $s=1$. For the case without axions, CP symmetry is violated and $\langle\hat{E}\rangle \neq 0$ for $\theta\neq0$, while introducing the axions leads to $\langle \hat{\theta}_\text{eff}\rangle =0$, restoring CP symmetry, as evident from $\langle\hat{E}\rangle = 0$. }
    \label{fig:CPEfield.}
\end{figure}

In QCD, the minimum of the vacuum energy lies at $\theta_\text{min}=0$, so introducing the axion field that drives $\langle\hat{\theta}_\text{eff}\rangle=0$ restores CP symmetry. Similarly, for $\theta \neq 0$, the Hamiltonian of the Schwinger model is not invariant under the CP transformation $\hat{E}\to -\hat{E}$, resulting in $\langle \hat{E}\rangle \neq0$ for any $\theta\neq 0$. After introducing the axion in the $s=1$ case, we found previously that the ground state satisfies $\langle\hat{\theta}_\text{eff}\rangle=0$, which implies a restoration of CP symmetry. This restoration induced by the introduction of the axion field in the model is shown in Fig.~\ref{fig:CPEfield.}. We observe that in the presence of the axion, $\langle \hat{E}\rangle$ is zero, which is required by CP conservation.

\begin{table}[ht]
\centering
\begin{tabular}{c|c|c|c}
$g^2$ & $\chi$ & $m_a$ & $\Delta E$  \\
\hline
$1$ & $0.0141$ & $0.0594$ & $0.0656$  \\
$5$ & $0.0888$ & $0.1489$ & $0.1486$\\
$10$ & $0.2114$ & $0.230$ & $0.230$ \\
\end{tabular}
\normalsize
\caption{The vacuum susceptibility $\chi$, the induced mass of the axion $m_a$, and the energy gap to the first excited state $\Delta E$ for various values of $g^2$. There is a constant approximate relation between $m_a$ and $\Delta E$, which improves as $g^2$ is increased.}
\label{tab:observables}
\end{table}

In effective descriptions of the axion in QCD, one introduces an axionic potential $V_\text{eff}(\theta_\text{eff})=\chi(1-\cos\theta_\text{eff})$, where $\chi$ is the topological susceptibility. This potential captures the effect of the $\theta$-dependent ground-state energy on the dynamics of the axion. Expanding the above expression for small $\theta_\text{eff}$ leads to a term $\frac{1}{2}m_a^2a^2(x)$, where the axion mass is given by $m_a=\frac{2\pi}{f_a}\sqrt{\chi}$. Similarly, we define the topological susceptibility $\chi$ in our model as the second derivative of the effective potential per axion site at its minimum, allowing us to extract the mass of the axion. Moreover, we can access the energy of the first excited state of the system using the MPS excitation ansatz~\cite{haegeman2012,osborne2024c}. This state is driven by axionic excitations, as indicated in the Supplemental Material~\cite{SM} by comparing observables between the first excited state and the ground state, showing that axionic observables have the most significant changes. As a result, the energy gap of the Hamiltonian is dominated by the mass of the axion with corrections suppressed by $1/g^2$. This energy gap is effectively independent of $\theta$ provided $\theta$ is not close to $\pm \pi$, which is the regime to which we restrict our discussion. The obtained results are shown in Table~\ref{tab:observables}. We observe that $m_a$ and the gap between the first excited state and the ground state, $\Delta E$, agree well with small deviations appearing at low $g^2$. This shows that in our model, the axion acquires its mass through the coupling to the gauge sector, in direct analogy with QCD.

\textbf{\textit{Summary.---}}
In this work, we studied ground-state properties of the Schwinger model in terms of the topological $\theta$-term using tensor network techniques. By introducing a scalar field that couples to the topological term, we investigated the response to a dynamical axion. We found that even in $1+1$D QED, under truncations on both the gauge fields and on the maximum axion number, the axion expectation value dynamically cancels the parameter $\theta$, as a result of the axion potential being minimized at zero. 

We found that for half-integer truncations of the electric field, the minimum of the axion potential deviates from $\theta=0$, resulting in an imperfect axion cancellation, especially at low values of the gauge coupling $g^2$. Still, it is expected that in the limit of $s\to \infty$ the behavior of both integer and half-integer truncations converges to a $2\pi$ periodic potential with minima at $\theta=0\;\text{mod}\;2\pi$. Furthermore, we analyzed how the axion restores CP symmetry, which is otherwise broken, resembling the way it resolves the strong CP problem in QCD. Finally, we verified that the generated mass from the axion potential is consistent with the energy gap to the first excited state, showing that even in the massive Schwinger model, several properties of the QCD axion can be reproduced.

\medskip
\footnotesize
\begin{acknowledgments}
\textbf{\textit{Acknowledgments.---}}
J.C.H.~acknowledges stimulating discussions with Takis Angelides, Lena Funcke, Karl Jansen, and Guo-Xian Su in the early stages of this project.
G.R., J.J.O., and J.C.H.~acknowledge funding by the Max Planck Society, the Deutsche Forschungsgemeinschaft (DFG, German Research Foundation) under Germany's Excellence Strategy – EXC-2111 – 390814868, and the European Research Council (ERC) under the European Union's Horizon Europe research and innovation program (Grant Agreement No.~101165667)—ERC Starting Grant QuSiGauge. N.B.~and T.M.~acknowledge support
by the DFG cluster of excellence ORIGINS funded by the Deutsche Forschungsgemeinschaft
(DFG) under Germany's Excellence Strategy - EXC-2094-390783311. N.B. acknowledges the European
Research Council advanced grant ERC-2023-ADG-Project EFT-XYZ. Views and opinions expressed are, however, those of the author(s) only and do not necessarily reflect those of the European Union or the European Research Council Executive Agency. Neither the European Union nor the granting authority can be held responsible for them. This work is part of the Quantum Computing for High-Energy Physics (QC4HEP) working group.
\end{acknowledgments}
\normalsize
\bibliographystyle{apsrev4-2}
\bibliography{biblio}
\clearpage
\onecolumngrid

\setcounter{equation}{0}
\setcounter{figure}{0}
\setcounter{table}{0}
\setcounter{page}{1}

\renewcommand{\theequation}{S\arabic{equation}}
\renewcommand{\thefigure}{S\arabic{figure}}
\renewcommand{\thetable}{S\arabic{table}}

\begin{center}
{\Large \textbf{Supplemental Material for ``Schwinger Model with a Dynamical Axion''}}\\
Gabriel Rouxinol, Tom Magorsch, Jesse J.~Osborne, Nora Brambilla, and Jad C.~Halimeh
\end{center}

\twocolumngrid
\section{Explicit breaking of the anomalous symmetry}
\begin{figure}[ht]
    \centering
    \includegraphics[width=1.0\linewidth]{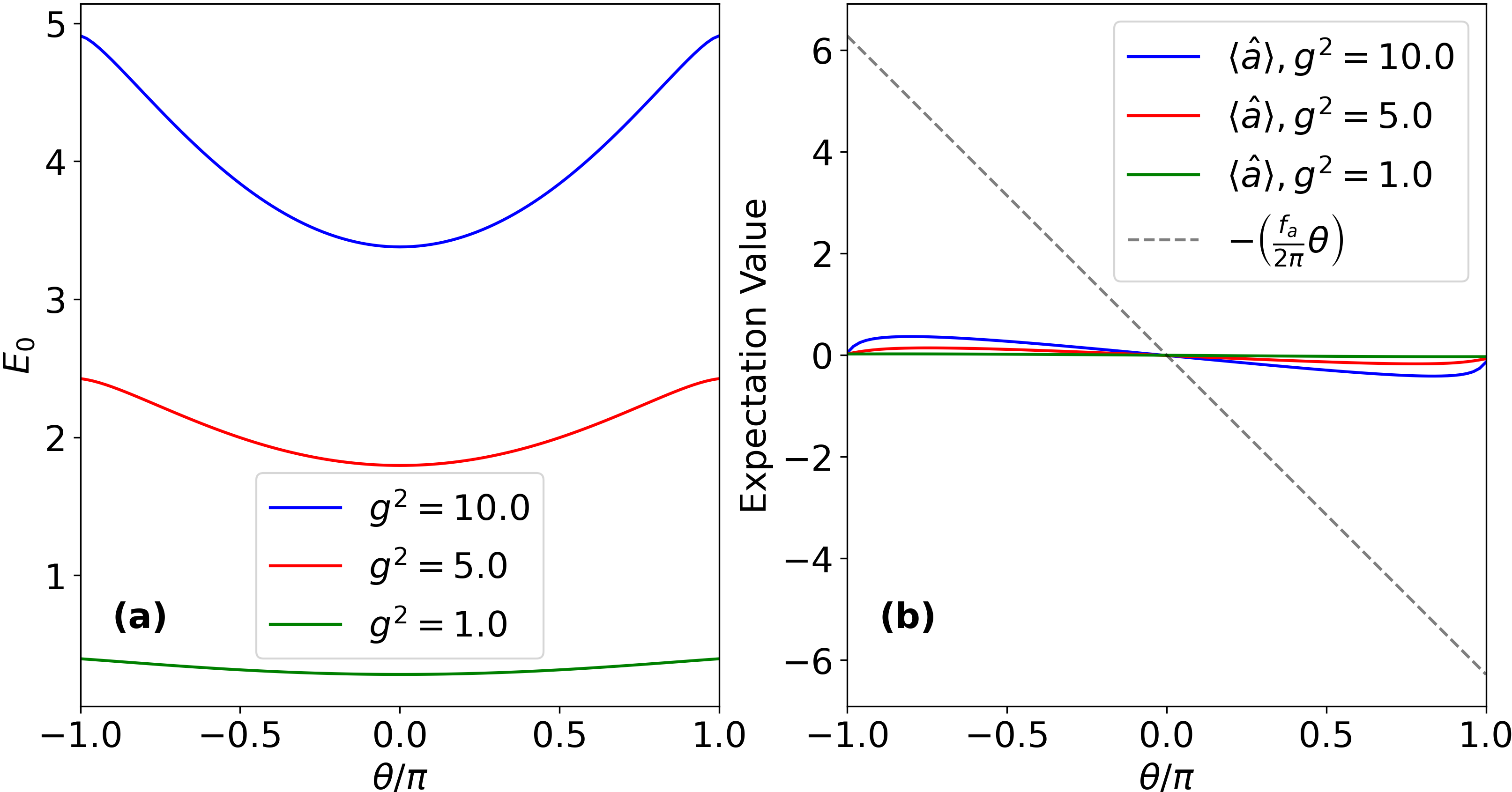}
    \caption{iDMRG calculations of (a) the ground-state energy of Hamiltonian~\eqref{eqn:FullAxionHamiltoinan} on the topological $\theta$-angle, (b) the expectation value of the axion field $\hat{a}$ as a function of $\theta$, for the spin $s=1$ QLM, $N_\text{max}=32$ and $m_a^2=0.5$. The ground-state energy is no longer a constant, and there is no axion cancellation, as a result of the breaking of the axion mechanism for a nonzero bare mass.}
    \label{fig:KineticSpic1Mass}
\end{figure}

We investigate the explicit breaking of the shift symmetry of the axion field by introducing a bare mass term $m_a^2\frac{\hat{a}_j^2}{2}$. Such an explicit breaking is expected to spoil axion cancellation and reintroduce a $\theta$-dependence into the ground-state energy. The Peccei-Quinn mechanism relies on an anomalous shift symmetry:
\begin{align}
    \theta &\to \theta -\alpha \nonumber \\
    \hat{a} &\to \hat{a}+ \frac{f_a}{2\pi} \alpha,
\end{align}
under which $\hat\theta_\text{eff}$ remains invariant, ensuring that the physics depends only on $\hat\theta_\text{eff}$. A bare axion mass term, however, explicitly breaks this symmetry. We demonstrate this in Fig.~\ref{fig:KineticSpic1Mass}, where we add a bare mass $m_a^2 = 0.5$ to the Hamiltonian. As shown in Fig.~\ref{fig:KineticSpic1Mass}(a), the ground-state energy now depends on 
$\theta$, and Fig.~\ref{fig:KineticSpic1Mass}(b) shows that the 
expectation value of the axion field no longer follows $-\frac{f_a \theta}{2\pi}$. The axion cancellation is spoiled.
\vspace{-7mm}
\section{Change of observables in the first excited state}
To verify that the first excitation is predominantly axionic and thereby justify our comparison between the energy gap and the axion mass, we study changes in the expectation values of three observables, one for each field type. We define $\Delta \hat{\mathcal{O}}$, as the difference of $\langle \hat{\mathcal{O}}\rangle$ between the first excited state and the ground state, for $\hat{\mathcal{O}} = \frac{1}{2}\hat{\psi}^\dagger_\ell \hat{\psi}_\ell, \hat{E}_{\ell, \ell+1}, \hat{n}_l$, where $\hat{n}_l=\hat{b}_l^\dagger \hat{b_l}$. The results are shown in Table \ref{tab:excitedobservables}. For $\Delta \hat{E}_{\ell, \ell+1}$ we report $0$, since the measured values oscillate around $0\pm 0.001$. For $\Delta \hat{n}_\ell$ with $g^2=1$, we report the value $3$ as the expectation value exhibits significant oscillations of the order $0.3-0.4$ around $3$ as $\theta$ is varied. As we can see, the most prevalent change is by far on the axion number, while the small change in $\frac{1}{2}\Delta\hat{\psi}^\dagger_\ell \hat{\psi}_\ell$, will be multiplied by the mass $m=0.25$, giving a contribution to the energy gap of around $0.005$ for $g^2=1,5$. Specifically for the case of $g^2=1$, this change is comparable to the total energy gap $\Delta E$, which justifies its discrepancy from $m_a$. Taken together with the agreement between the energy gap $\Delta E$ and the axion mass at larger $g^2$, indicates that the first excitation of the system is driven mostly by axion fluctuations. 

\begin{table}[ht]
\centering
\begin{tabular}{c|c|c|c}
$g^2$ & $\frac{1}{2}\Delta\hat{\psi}^\dagger_\ell \hat{\psi}_\ell$ & $\Delta \hat{E}_{\ell, \ell+1}$ & $\Delta \hat{n}_l$  \\
\hline
$1$ & $0.02$ & $0$ & $3$  \\
$5$ & $0.02$ & $0$ & $1.68$\\
$10$ & $0.01$ & $0$ & $1.14$ \\
\end{tabular}
\normalsize
\caption{The change in the expected value of the observables $\hat{\mathcal{O}} = \overline{\psi}_\ell\psi_\ell, \hat{E}_{\ell, \ell+1}, \hat{n}_l$, for different values of $g^2$. The first excited state is dominated by axionic excitations, allowing us to compare the energy gap with the axion mass.}
\label{tab:excitedobservables}
\end{table}
\vspace{-7mm}
\section{Impact of truncations on Results}
\begin{figure}[H]
    \centering
    \includegraphics[width=1.0\linewidth]{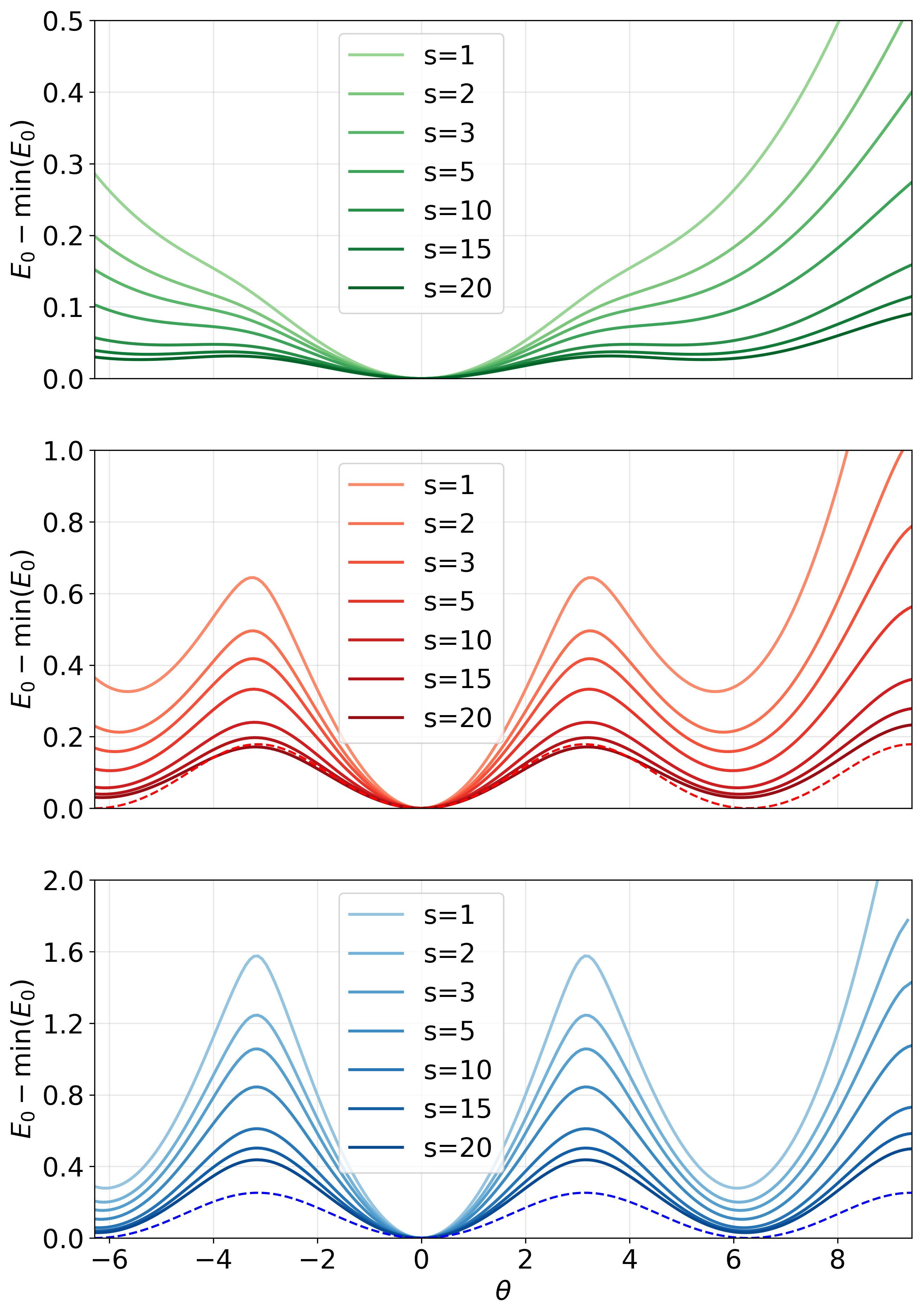}
    \caption{iDMRG calculations of the ground state dependence of Hamiltonian~\eqref{eqn:FullAxionHamiltoinan} on the topological $\theta$-angle at $g^2=1$ (green), $g^2=5$ (red) and $g^2=10$ (blue) $s\in \{1, 2, 3, 5, 10,15,20\}$, with increasing $s$ going from light to dark shades in the curves. The dashed line represents the expected behavior in the continuum calculated for a small fermion mass~\cite{colemanMoreMassiveSchwinger1976}.}
    \label{fig:higherspincheck}
\end{figure}

\begin{figure}[H]
    \centering
    \includegraphics[width=1.0\linewidth]{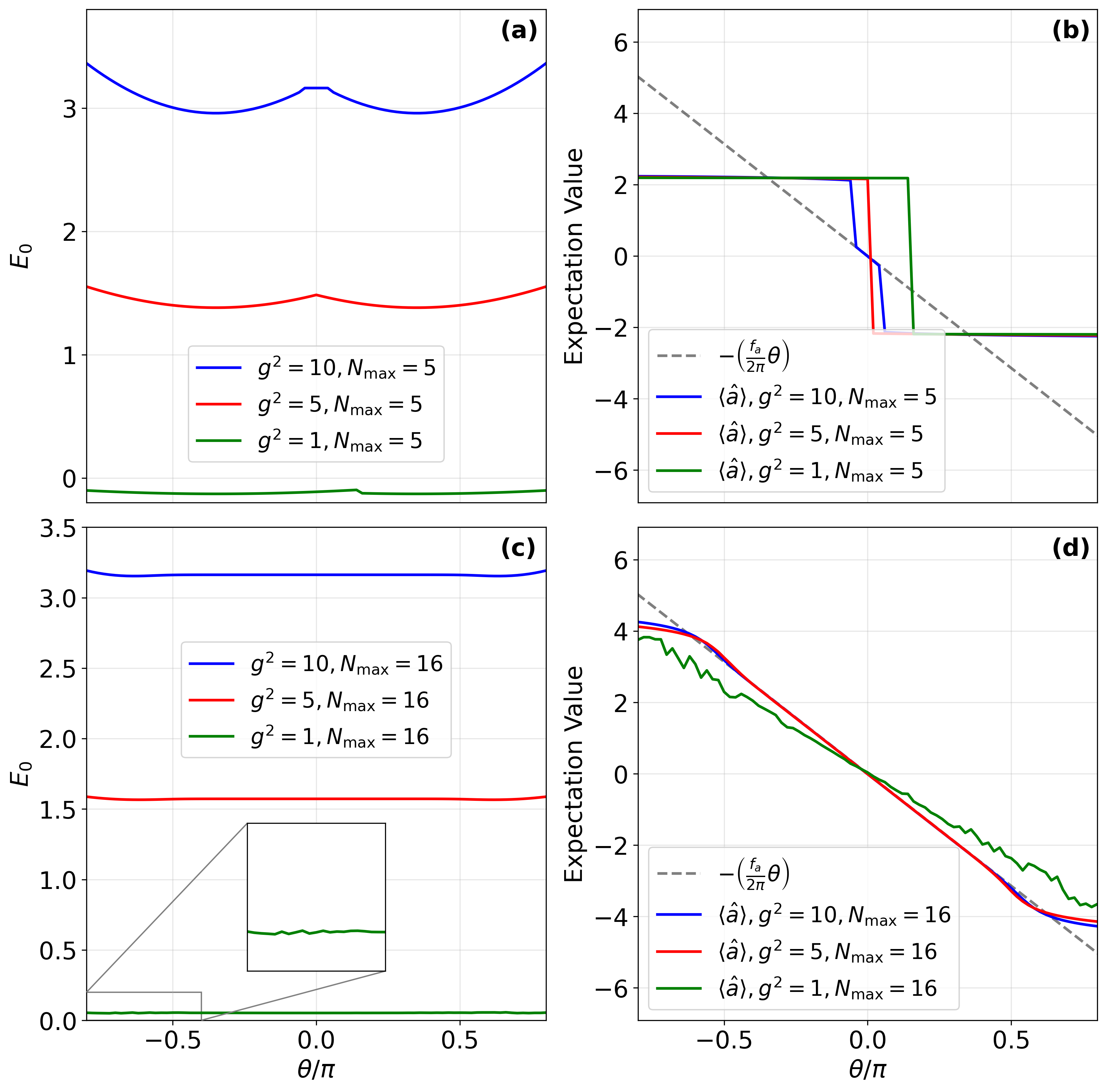}
    \caption{iDMRG calculations of the ground-state energy of Hamiltonian~\eqref{eqn:FullAxionHamiltoinan} and the expectation value of the axion field $\hat{a}$ as a function of the topological $\theta$-angle for, respectively, (a), (b) $N_\text{max}=5$ and (c), (d) $N_\text{max}=16$, for the spin $s=1$ QLM. For $N_\text{max}=5$, only for $g^2=10$ and $\theta$ close to zero, the ground-state energy is independent of the angle. Axion cancellation only is observed in that region. For $N_\text{max}=16$ and $g^2=5, 10$, the energy is independent of $\theta$, and axion cancellation happens, albeit in a smaller region than for $N_\text{max}=32$. However, for $g^2=1$, the truncation is too small to see the desired effects.}
    \label{fig:KineticSpic1N5}
\end{figure}

Here we study the ground-state energy of the Schwinger model for the values of $g^2 \in \{1,5,10\}$ used throughout this work, considering spin truncations up to $s=20$. This shows that although the chosen parameters are not close to the continuum limit of the Schwinger model, taking the Kogut--Susskind limit of the QLM allows us to recover the $2\pi$ periodicity of $E_0$, as shown in Fig.~\ref{fig:higherspincheck}. For $g^2=5, 10$, we include as dashed line the continuum expectation computed in~\cite{colemanMoreMassiveSchwinger1976} for small fermion mass, which was compared to the lattice Schwinger model using tensor network methods in~\cite{PhysRevD.101.054507}. This confirms that, despite being far from the continuum limit, our model captures the qualitative behavior expected from the quantum field theory in the continuum. Nevertheless, extracted quantities such as the topological susceptibility do not follow their continuum scaling. In particular, in the regime  $\frac{m}{g} \ll 1$, the continuum theory predicts $\chi \propto m g$, which is not observed in Table~\ref{tab:observables}. Furthermore, to investigate the dependence on the truncation of the bosonic per-site Hilbert space of the axion, we repeat the ground state analysis for $N_\text{max}=5$  and for $N_\text{max}=16$ in Fig.~\ref{fig:KineticSpic1N5}. For $N_\text{max}=5$, we find that the flatness of the energy dependence on $\theta$ is lost, except for $g^2=10$, where the ground-state energy becomes flat for a narrow interval around $\theta=0$. Within this interval, we can still confirm the cancellation of $\langle \hat{a}\rangle$. On the other hand, as we increase $N_\text{max}$ to $16$, the results are very similar to those for $N_\text{max}=32$ at $g^2=5, 10$, although the interval over which $\langle \hat{a}\rangle$ exhibits the expected cancellation is reduced. For $g^2=1$, however, the axion expectation value no longer matches what would be expected in the case of axion cancellation, even though the ground-state energy only has a weak dependence on the angle $\theta$. This can be understood as a result of $\frac{g^2}{f_a} \ll 1$, which leads to changes in the expected value of the axion having low influence on the energy of the ground state. 

\end{document}